%% file: proceedingv2.tex
\documentclass{PoS}
\pdfoutput=1 
\usepackage{amsmath}
\usepackage{amssymb}
\usepackage{graphicx}
\usepackage{paralist}
\usepackage{subfigure}
\usepackage[normalem]{ulem}
\usepackage{cancel}

\def\beq{\begin{equation}}
\def\eeq{\end{equation}}
\def\bea{\begin{eqnarray}}
\def\eea{\end{eqnarray}}

\include{iQ_newco}

\newcommand{\at}{\makeatletter @\makeatother}

\title{\vspace{-0.5cm}On the intrinsic heavy quark content of the nucleon}

\ShortTitle{On the intrinsic bottom content of the nucleon and its impact at the LHC}

\author{\speaker{Florian Lyonnet}$^a$, Aleksander Kusina$^b$, Karol Kova\v{r}\'{\i}k$^c$,Tom\'{a}\v{s} Je\v{z}o$^d$, Fred Olness$^a$, Ingo Schienbein$^b$ and Ji-Young\ Yu$^a$ \\\\
\llap{$^a$}Southern Methodist University, Dallas, TX 75275, USA\\
\llap{$^b$}Laboratoire de Physique Subatomique et de Cosmologie,\\
				Universit\'e Joseph Fourier/CNRS-IN2P3/ INPG,\\
				53 Avenue des Martyrs, F-38026 Grenoble, France\\
\llap{$^c$}Institut f{\"u}r Theoretische Physik, Westf{\"a}lische Wilhelms-Universit{\"a}t M{\"u}nster,\\
        Wilhelm-Klemm-Stra{\ss}e 9, D-48149 M{\"u}nster, Germany\\
\llap{$^d$}Universit\`a di Milano-Bicocca and INFN, Sezione di Milano-Bicocca,\\
			  Piazza della Scienza 3, 20126 Milano, Italy\\
				E-mail: \email{flyonnet@mail.smu.edu}, \email{kusina@lpsc.in2p3.fr}, \email{kovarik@particle.uni-karlsruhe.de}, \email{tomas.jezo@mib.infn.it}, \email{olness@mail.smu.edu}, \email{schien@lpsc.in2p3.fr}, \email{yu@lpsc.in2p3.fr}}

\abstract{

We demonstrate that to a very good approximation the scale-evolution
of the intrinsic heavy quark content of the nucleon is governed by
non-singlet evolution equations. This allows us to analyze the intrinsic heavy
quark distributions without having to resort to a full-fledged global analysis
of parton distribution functions. This freedom is then exploited to model intrinsic
bottom distributions which are so far missing in the literature in order to 
estimate the impact of this non-perturbative contribution to the bottom-quark
PDF, and on parton--parton luminosities at the LHC.
This technique can be applied to the case of intrinsic charm, 
albeit within the limitations outlined in the following. 
}

\FullConference{The XXIII International Workshop on Deep Inelastic Scattering and Related Subjects\\
		April 27 - May 1, 2015\\
                Southern Methodist University\\
		Dallas, Texas 75275}

\begin{document} 
\maketitle
\section{Introduction}

Heavy quark parton distribution functions (PDFs) play an important role
in several Standard Model (SM) and New Physics (NP) processes at the
CERN Large Hadron Collider (LHC), e.g. see~\cite{Maltoni:2012pa} for key processes involving the bottom quark PDF. 
In the standard approach, the heavy quark distributions are generated {\em radiatively}, according to DGLAP evolution equations starting with a perturbatively calculable boundary condition. However, a purely perturbative, {\it extrinsic}, treatment where the heavy quarks are radiatively generated might {\em not} be adequate and there are a number of models that postulate a non-perturbative, {\it intrinsic}\footnote{Note that in our definition the intrinsic PDF is a twist 2 object.}, heavy quark component e.g. light-cone~\cite{Brodsky:1980pb,Brodsky:1981se} and meson cloud models~\cite{Navarra:1995rq,Paiva:1996dd,Melnitchouk:1997ig}.

Along the years, different groups have estimated the amount of intrinsic charm (IC) allowed in the nucleon by performing global fits, see~\cite{Brodsky:2015fna} for a recent review. Among them, the two most recent analyses~\cite{Gao:2013xoa,Jimenez-Delgado:2014zga} set significantly different limits on the allowed IC contribution highlighting the utility of the techniques discussed in this paper as we can freely adjust the amount of IC/IB contributions without having to regenerate a complete global analysis for each case. 

In this contribution, we summarize a technique which can provide IB PDFs for any generic
non-IB PDF set, that we have introduced in~\cite{Lyonnet:2015dca}. Our approach exploits
the fact that the intrinsic bottom PDF evolves (to an excellent precision) according to
a standalone non-singlet evolution equation; that allows us to easily obtain a matched set
of IB and non-IB PDFs. Note that because existing data entering global analyses of proton
PDFs do not constrain the IB PDF, it would not be useful to try and obtain information on
the IB content of the nucleon using a global fit.

The rest of this paper is organized as follows.
In Sec.~\ref{sec:intrinsic}, we demonstrate that to a good approximation the scale-evolution
of the intrinsic PDF is governed by a non-singlet evolution equation, and construct a set of matched IC/IB PDFs.
In Sec.~\ref{sec:numerics}, we use the IB PDFs to obtain predictions
for parton--parton luminosities relevant at the LHC.
Finally, in Sec.~\ref{sec:conclusions}, we summarize our results and
present conclusions.

\section{Intrinsic heavy quark PDFs} 
\label{sec:intrinsic}

\subsection{Definition and Evolution}
\label{sec:definition}
In the context of a global analysis of PDFs the different parton flavors are specified via a boundary condition at the input scale $\mu_0$ which is typically of the order ${\cal O}(1\ \GeV)$. Solving the DGLAP evolution equations with these boundary conditions allows us to determine the PDFs at higher scales $\mu > \mu_0$. A non-perturbative (intrinsic) heavy quark distribution $Q_1$ can then be defined at the input scale $\mu_0$ as the difference of the full boundary condition for the heavy quark PDF $Q$ and the perturbatively calculable (extrinsic) boundary condition $Q_0$.

Denoting the vector of light quarks as `$q$' and the heavy quark distribution by `$Q$' (where $Q=c$ or $Q=b$) the  Dokshitzer-Gribov-Lipatov-Altarelli-Parisi (DGLAP) evolution equations read 
\begin{eqnarray}
\label{eq:DGLAP2a}
\dot g &=& P_{gg}\otimes g+P_{gq}\otimes q+P_{gQ}\otimes Q_0+ {\cancel{P_{gQ}\otimes Q_1}}\, ,
 \\
\label{eq:DGLAP2b}
\dot q &=& P_{qg}\otimes g+P_{qq}\otimes q+P_{qQ}\otimes Q_0+ {\cancel{P_{qQ}\otimes Q_1}}\, ,
 \\
\label{eq:DGLAP2c}
\dot Q_0 + \dot Q_1 &=& P_{Qg}\otimes g+P_{Qq}\otimes q+P_{QQ}\otimes Q_0+ P_{QQ}\otimes Q_1 \, .
\end{eqnarray}
In which we substituted $Q=Q_0+Q_1$ where $Q_0$ denotes the usual radiatively generated
extrinsic heavy quark component and $Q_1$ is the non-perturbative intrinsic heavy quark
distribution.\footnote{Strictly speaking, the decomposition of $Q$ into $Q_0$ and $Q_1$ is 
defined at the input scale where the calculable boundary condition for $Q_0$ is known. 
Consequently, $Q_1:=Q-Q_0$ is known as well.
Only due to the approximations in Eqs.\ \protect\eqref{eq:DGLAP2a} and \protect\eqref{eq:DGLAP2b}
it is possible to entirely decouple $Q_0$ from $Q_1$ so that the decomposition becomes meaningful
at any scale.}

Neglecting the crossed out terms which give a tiny contribution 
to the evolution of the gluon and light quark distributions the
system of evolution equations can be separated into two independent parts.
For the system of gluon, light quarks and extrinsic heavy quark ($g,q,Q_0$) one recovers the same evolution equations as in the standard approach without an intrinsic heavy quark component. For the intrinsic heavy quark distribution, $Q_1$, one finds a standalone non-singlet evolution equation, $\dot Q_1 = P_{QQ}\otimes Q_1 \, .$ 

Allowing for a small violation of the sum rule it is possible to entirely decouple the analysis of the intrinsic heavy quark distribution from the rest of the system. The PDFs for the gluon, the light quarks and the extrinsic heavy quark can be taken from a global analysis in the standard approach where they already saturate the momentum sum rule. On top of these PDFs the intrinsic heavy quark PDF can be determined in a standalone analysis using the non-singlet evolution equation. 

This induces a violation of the momentum sum rule by the term $\int_0^1\ \der x\  x\  \left(Q_1 + \bar{Q}_1\right)$ which, however, is very small for bottom quarks.\footnote{It is also acceptable in case of charm provided that the allowed normalization of IC is not too big.} We will perform numerical checks of the validity of our approximations below. 

\subsection{Modeling the boundary condition}
\label{sec:bc}

The BHPS model~\cite{Brodsky:1980pb} predicts the following $x$-dependence
for the intrinsic charm (IC) parton distribution function:
\begin{equation}
c_1(x) = \bar c_1(x) \propto x^2 [6 x (1+x) \ln x + (1-x)(1+10 x+x^2)]\, .
\label{eq:bhps}
\end{equation}
We expect the $x$-shape of the intrinsic bottom distribution $b_1(x)$ to be very similar to the one of the intrinsic charm distribution. Furthermore, the normalization of IB is expected to be parametrically suppressed with respect to IC by a factor $m_c^2/m_b^2 \simeq 0.1$. To fix the freedom related to the scale of the boundary condition we use in the following the {\it Same Scales} boundary condition, which remains valid at any scale $Q$: $b_1(x,m_c) = \frac{m_c^2}{m_b^2} c_1(x,m_c)\, .$ Finally, let us note that it would be no problem to work with asymmetric boundary conditions, $\bar c_1(x) \ne c_1(x)$
and $\bar b_1(x) \ne b_1(x)$, as predicted for example by meson cloud models~\cite{Paiva:1996dd}.

\subsection{Intrinsic heavy quark PDFs from non-singlet evolution}
\label{sec:PDFdef}

Eq.~\eqref{eq:bhps} is used to define the initial $x$-dependence at the scale of the charm mass, the normalization is fixed to match the one predicted by the CTEQ6.6c0 fit. The IB PDF was generated using the {\it Same Scales} boundary conditions (see above) together with the same $x$-dependent input of Eq.~\eqref{eq:bhps}. Both PDFs were then evolved according to the non-singlet evolution equation and the corresponding grids were produced. Note that because in our approximation, the evolution of the intrinsic charm and bottom PDFs is completely decoupled; the normalization of our PDFs can be easily changed by means of simple rescaling.

\subsection{Numerical validation}
\label{sec:tests}

In order to test the ideas presented above, we use the CTEQ6.6c series of intrinsic charm fits and in particular CTEQ6.6c0 and CTEQ6.6c1, which employ the BHPS model with $1\%$ and $3.5\%$ IC probability, respectively.\footnote{This corresponds to the values of 0.01 and 0.035 of the first moment of the charm PDF, $\int dx  \, c(x)$, calculated at the input scale $Q_0=m_c=1.3\ \GeV$.} In the following we compare our approximate IC PDFs supplemented with the central CTEQ6.6 fit, which has a radiatively generated charm distribution, with the CTEQ6.6c0 and CTEQ6.6c1 sets where IC has been obtained from global analysis.

\begin{figure}
\begin{center}
\subfigure[]{
\label{fig:testa}
\includegraphics[angle=0,width=0.48\textwidth]{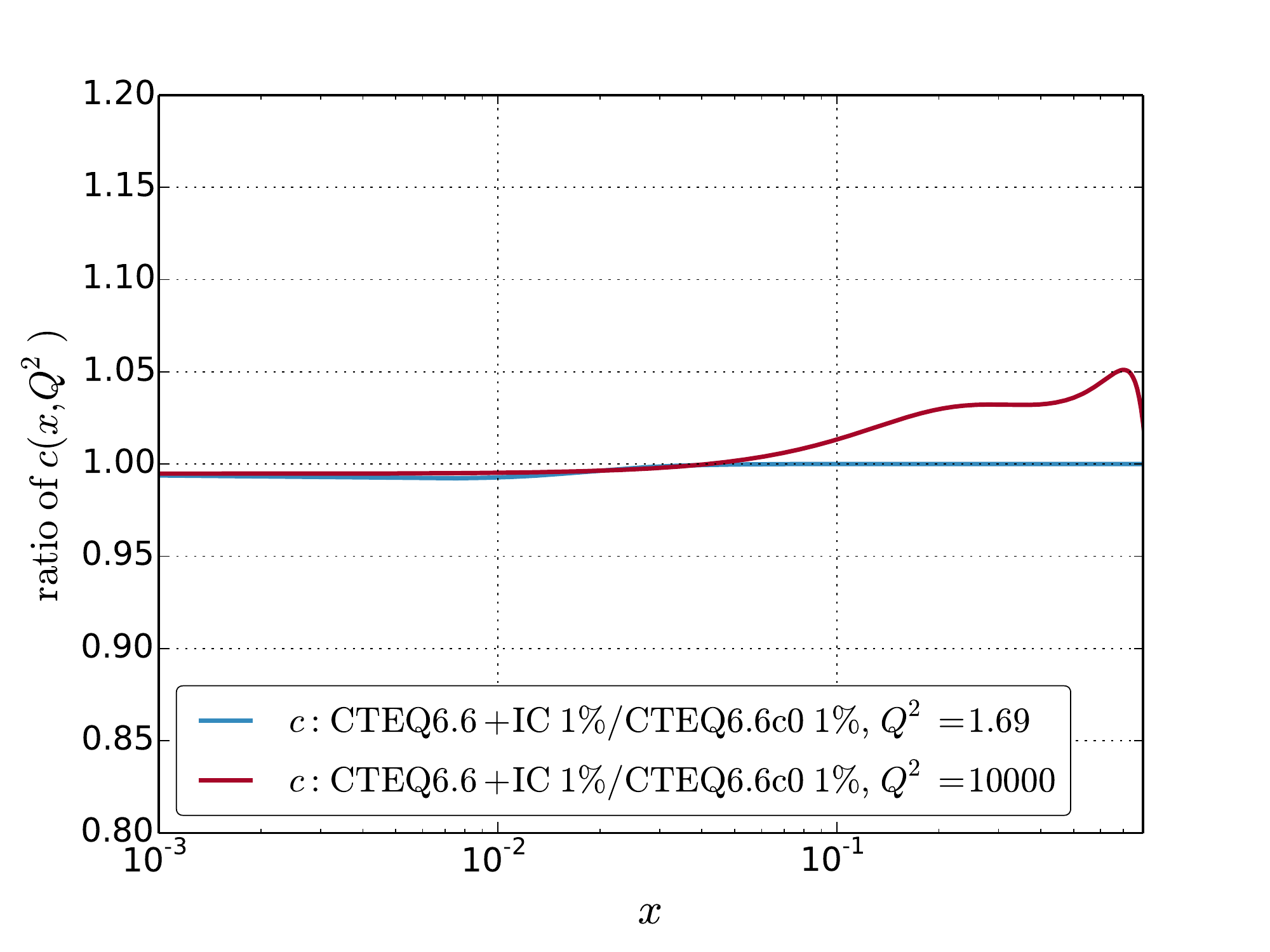}}
\subfigure[]{
\label{fig:testb}
\includegraphics[angle=0,width=0.48\textwidth]{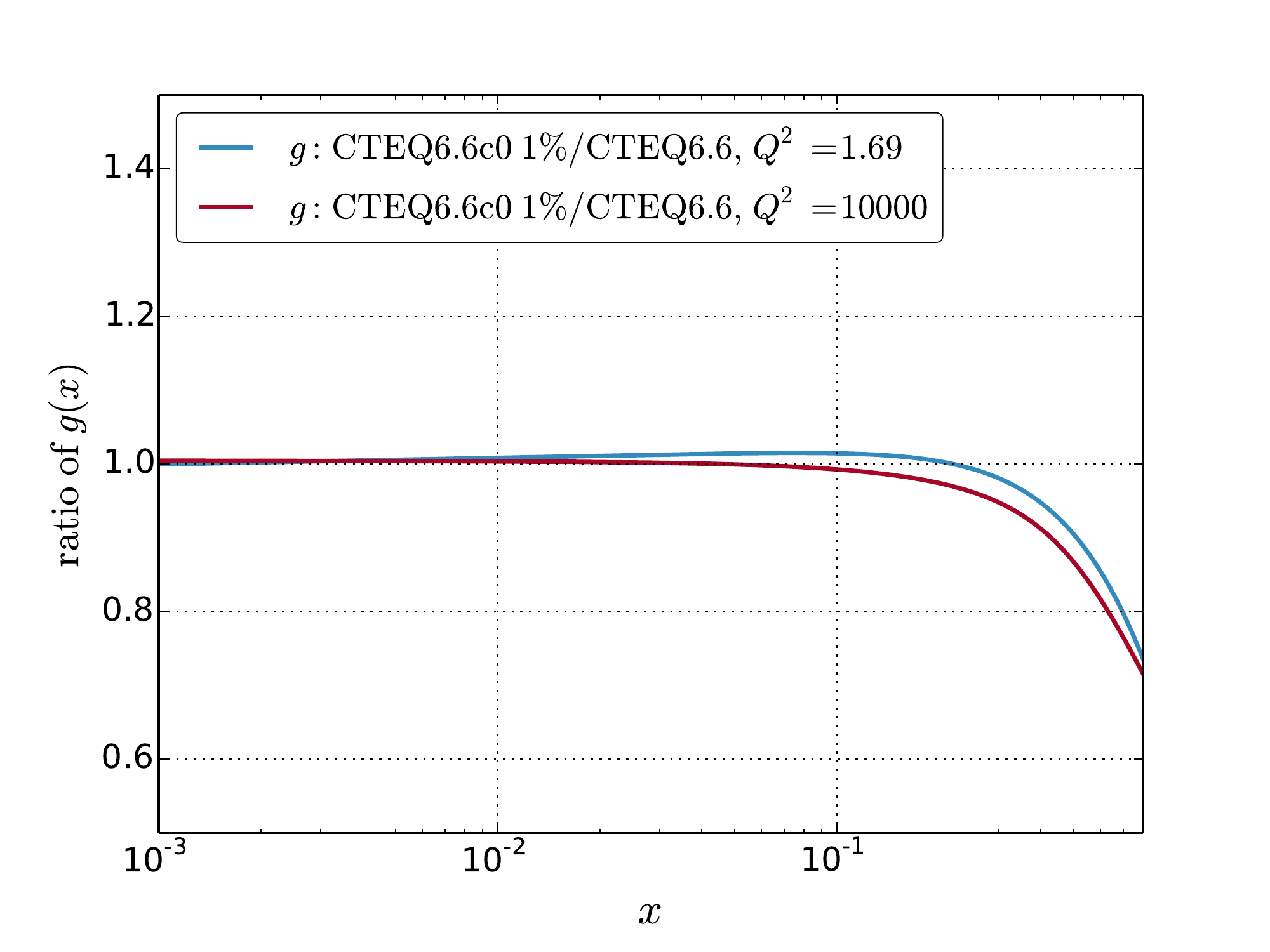}}
\caption{(a) Ratio of the CTEQ6.6 + IC 1\% and CTEQ6.6c0 (1\%) charm distributions. (b) Ratio of the CTEQ6.6c0 and the CTEQ6.6 gluon distributions. The results are shown as function of $x$ for two scales, $Q^2=1.69$ and $Q^2=10000\ \GeV^2$.}
\end{center}
\end{figure}

In Fig.~\ref{fig:testa}, one can see that the difference between the sum $c_0+c_1$ and the CTEQ6.6c0 charm distribution is tiny at low $Q^2$, and smaller than $5\%$ at the higher $Q^2$. 

The inclusion of the intrinsic charm distribution will alter the other parton distributions, most notably the gluon PDF. In order to gauge this effect we compare in Fig.~\ref{fig:testb} the gluon distribution from the CTEQ6.6c0 analysis with the one from the standard CTEQ6.6 fit. For small $x$ ($x<0.1$) the gluon PDF is not affected by the presence of a BHPS-like intrinsic charm component which is concentrated at large $x$. We note that at large-$x$, where most of the difference lies, the gluon distribution is already quite small and the uncertainty of the gluon PDF is sizable (of order of 40 -- 50\% for the CTEQ6.6 set).

We conclude that for most applications,  adding a standalone intrinsic charm distribution to an existing standard  global analysis of PDFs is internally consistent and leads only to a small error. Moreover, for the case of intrinsic bottom which is additionally suppressed, the accuracy of the approximation will be even better. For completeness we also provided similar validation using the parton--parton luminosities in~\cite{Lyonnet:2015dca} to which we refer the reader for further details.

\section{Possible effects of IC/IB on LHC observables}
\label{sec:numerics}

We study the effects of both IC and IB on parton--parton luminosities,
$\frac{d\mathcal{L}_{ij}}{d\tau}$, at LHC with $\sqrt{S}=14$ TeV.
We use the following definition of luminosity
\begin{equation}
\frac{d\mathcal{L}_{ij}}{d\tau}(\tau,\mu) = \frac{1}{1+\delta_{ij}}\frac{1}{\sqrt{S}}\int_\tau^1 \frac{dx}{x}
                       \Big[ f_i(x,\mu)f_j(\tau/x,\mu) + f_j(x,\mu)f_i(\tau/x,\mu) \Big],
\label{eq:pplumi}
\end{equation}
where $\tau=x_1 x_2$ and $f_i$ are parton distribution functions.
This allows us to assess the relevance of a non-perturbative heavy quark component for the production of new heavy particles coupling to the SM fermions.

%

\begin{figure}[t]
	\centering
	\includegraphics[width=0.45\textwidth]{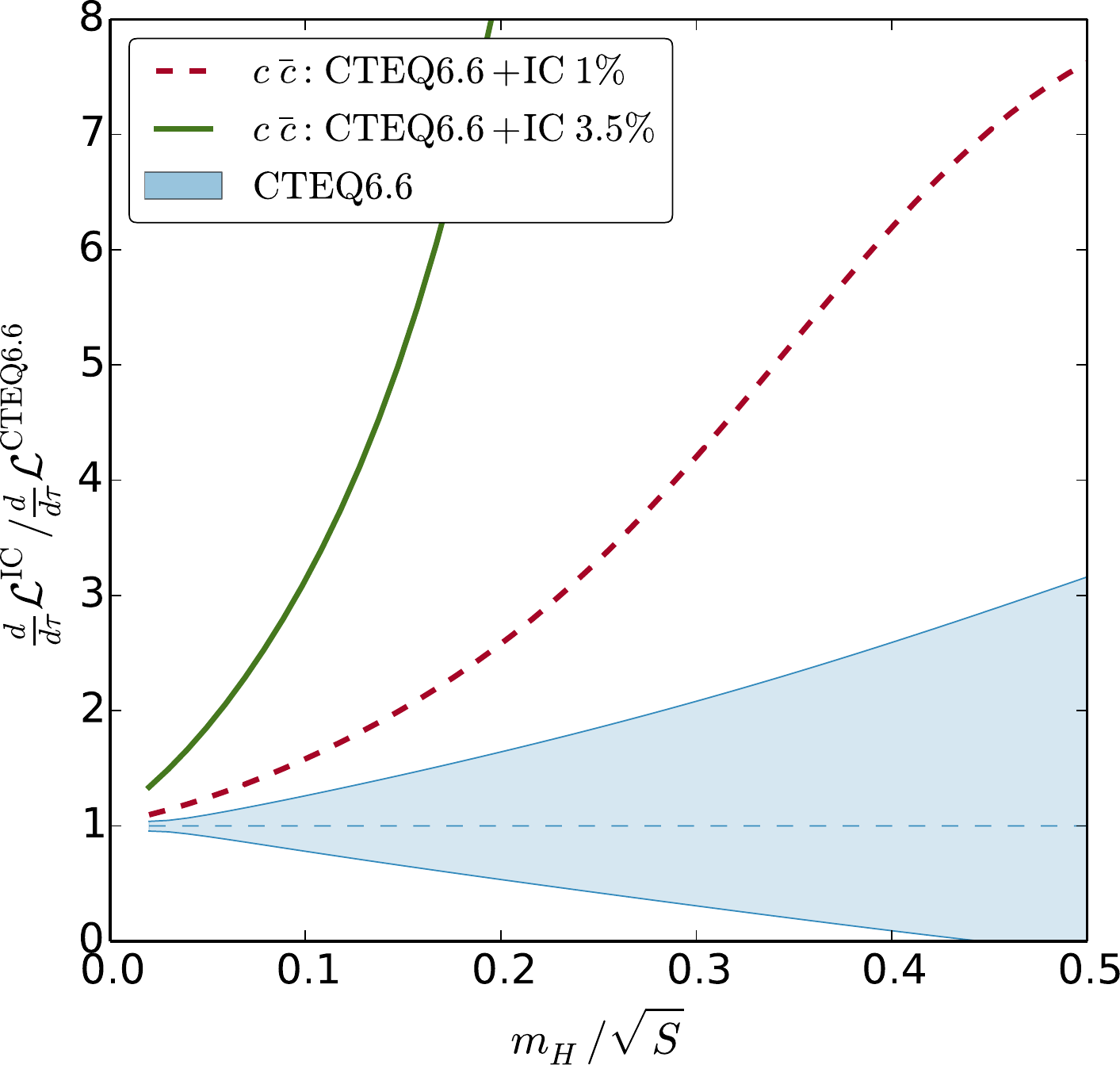}
	\includegraphics[width=0.45\textwidth]{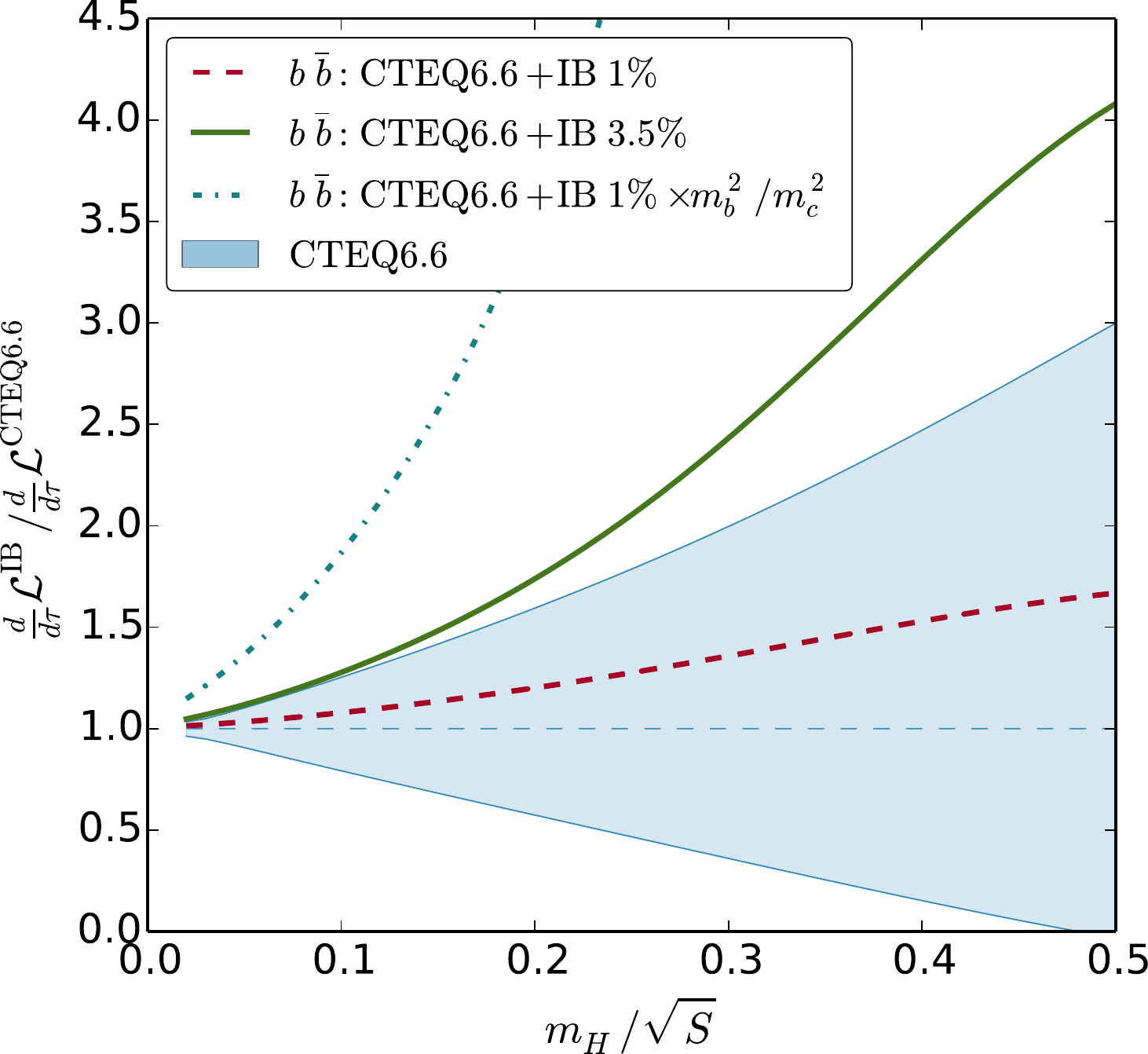}
	\caption{Ratio of $c \bar c$ luminosities (left) and $b \bar b$ luminosities (right) at the LHC14 
	for charm(bottom)-quark PDF sets with and without an intrinsic component as a function of $\sqrt{\tau}=m_H/\sqrt{S}$. 
	The ratio for the $c \bar c$ ($b \bar b)$ luminosity (solid, green line) in the left (right) figure reaches values of 50 (17) at $\sqrt{\tau}=0.5$. In addition to the curves with 1\% normalization (red, dashed lines) we include the results for the 3.5\% normalization (green, solid lines)
which was found to be still compatible with the current data~\cite{Nadolsky:2008zw}.
	}
	\label{fig:ratio_w_uncertainty_c}
\end{figure}

To explore how the presence of IC and IB would affect physics observables with a non-negligible heavy quark initiated subprocesses, in Fig.~\ref{fig:ratio_w_uncertainty_c} we show the ratios of luminosities for charm and bottom with and without an intrinsic contribution for 1\% and 3.5\% normalizations. Furthermore, since there are no experimental constraints on the IB normalization, in Fig.~\ref{fig:ratio_w_uncertainty_c} (right), we also include an extreme scenario where we remove the usual  $m_c^2/m_b^2$ factor; thus, the first moment of the IB is 1\% at the initial scale $m_c$.
As expected, in the case of IB the effect is smaller but for the $b \bar{b}$ luminosity  the
IB with 3.5\% normalization leads to a curve which lies clearly above 
the error band of the purely perturbative result. 
In the extreme scenario (which is not likely but by no means excluded) the IB component has a big effect 
on the $b\bar{b}$. 

Note that similar results for the $c g$ and $b g$ luminosities have been obtained but are not shown here, see~\cite{Lyonnet:2015dca}.

\section{Conclusions}
\label{sec:conclusions}

In this article, we presented a method to generate a matched IC/IB distributions
for any PDF set without the need for a complete global re-analysis. This allows one to easily carry
out a consistent analysis including intrinsic heavy quark effects. 
Because the evolution equation for the intrinsic heavy quarks decouples, 
we can freely adjust  the normalization of the IC/IB PDFs.

For the IB, our approximation holds to a very good precision.
For the IC, the error increases (because the IC increases), yet our method is still useful.
For an IC normalization of  1-2\%, the error is less than the  PDF uncertainties at the large-$x$ where the IC is relevant.
For a larger normalization, although the error may be the same order as the  PDF uncertainties, 
the IC effects also grow and can be separately distinguished from the case without IC. In any case, the IC/IB represents a non-perturbative systematic effect 
which should be taken into account.

The method presented here greatly simplifies our ability to estimate the impact of the intrinsic heavy quark effects on the new physics searches. It can also be very useful in searching for and constraining the intrinsic charm and bottom components of the nucleon by itself. In particular in the future facilities such as an Electron Ion Collider (EIC), the Large Hadron-Electron collider (LHeC), or AFTER\at LHC.

The PDF sets for intrinsic charm and intrinsic bottom discussed in this analysis
(1\% IC, 3.5\% IC, 1\% IB, 3.5\% IB) are available from the authors upon request.

\bibliographystyle{JHEP}
\bibliography{proceeding2.bbl}

\end{document}

%% file: iQ_newco.tex
\newcommand{\GeV}{ {\rm GeV}}

\newcommand{\der}{\ensuremath{{\operatorname{d}}}}